\documentclass[a4paper,11pt]{article}
\pdfoutput=1 

\usepackage{jheppub}

\usepackage[T1]{fontenc}
\usepackage[svgnames]{xcolor}
\usepackage{parskip}
\usepackage{breqn}
\usepackage{comment}
\usepackage{float}
\allowdisplaybreaks

\usepackage{slashed}
\usepackage{bm}

\newcommand{\be}{\begin{equation}\begin{aligned}}
\newcommand{\ee}{\end{aligned}\end{equation}}
\newcommand{\nn}{\nonumber}

\title{\boldmath Perturbative K\"ahler Moduli Inflation}

\author[a]{Mishaal Hai,}
\author[b,c]{Ahmed Rakin Kamal,}
\author[a]{Noshin Ferdous Shamma,}
\author[d]{Md Shaikot Jahan Shuvo}

\affiliation[a]{\small Dipartimento di Fisica e Astronomia, Universit\`a di Bologna, via Irnerio 46, 40126 Bologna, Italy}
\affiliation[b]{Department of Theoretical Physics and Astrophysics, Faculty of Science,
Masaryk University, Kotl\'a\v{r}sk\'a 2, CS-61137 Brno, Czechia}
\affiliation[c]{\small Department of Mathematics and Natural Sciences,
BRAC University, 66 Mohakhali, Dhaka 1212, Bangladesh}
\affiliation[d]{\small Initiative for the Theoretical Sciences, The Graduate Center, CUNY,
365 Fifth Ave, New York, NY 10016, USA}

\emailAdd{mishaalhai97@gmail.com}
\emailAdd{ahmedrakinkamaltunok@gmail.com}
\emailAdd{shammanoshin@gmail.com}
\emailAdd{sjshuvo70@gmail.com}

\abstract{In this work, we present two classes of inflationary models in the framework of type IIB string theory. The inflatons correspond to blow-up K\"ahler modulus arising from compactifying type IIB string theory on a Calabi-Yau. Using perturbative corrections, we first highlight a procedure for stabilising more than one K\"ahler modulus. For the case of two K\"ahler moduli, we explicitly construct two classes of inflationary potentials within the K\"ahler cone which satisfy both EFT and cosmological constraints. The first class of models, arising from moduli redefinition of the blow-up mode, garners a potential of the form $V(\phi)=V_{0}(1+C_{1} \phi^{2/3})$ align with CMB data with tensor-to-scalar ratio $r\lesssim 10^{-2}$. The second class of models, which have been recently proposed as loop blow-up inflation, have a form $V(\phi)=V_{0}(1+C_{2}\phi^{-2/3})$, also agrees with CMB data with tensor-to-scalar ratio $r\lesssim 10^{-8}$. Our work differs from the original loop blow-up inflation in terms of stabilization mechanism and subsequently the tensor-to-scalar ratio.

}

\begin{document} 
\maketitle
\flushbottom

\section{Introduction}
From observations, we know that the universe is going through an accelerated expansion (and probably went through one in the very early universe). To explain the current stage of the universe, we describe it by virtue of a positive cosmological constant. However, constructing a dS solution from string theory is a difficult endeavour due to no-go theorem \cite{Maldacena:2000mw} and Swampland conjectures \cite{Obied:2018sgi}. The Maldacena-Nunez no-go theorem forbids the construction of a warped de-Sitter solution using classical supergravity ingredients. However, stringy ingredients like orientifolds can be used to evade the no-go theorem as done in \cite{Giddings:2001yu} which is still being actively studied \cite{Frey:2025rvf}. Moreover, several interesting ways to get around the no-go theorem has been developed in recent years \cite{Burgess:2024jkx, Andriot:2025cyi, Tringas:2025uyg}. A point to note that even if an evolving dark energy is preferred by observations, an embedding in string theory proves to be difficult \cite{Cicoli:2018kdo}, however, a positive story can still be told \cite{Cicoli:2024yqh, Dutta:2021bih}. 

When a type IIB string is compactified on a 6d Calabi-Yau orientifold, the resulting theory in 4 dimensions is a $\mathcal{N}=1$ supergravity theory. This theory admits a plethora of scalars that need to have a potential in order to have realistic phenomenology \cite{Quevedo:2014xia, McAllister:2023vgy, Cicoli:2023opf}.\footnote{\cite{Banks:2025nfe} raised an issue of parameters in string theory not being vacuum expectation value of moduli fields which was clarified in \cite{Sen:2025bmj, Sen:2025ljz}.}$\mathcal{N}=1$ supergravity can be written in terms of the superpotential $W$ and the K\"ahler potential $K$. The story of the scalars is understood by looking at the F-term scalar potential. The superpotential $W$ depends on the complex structure moduli, the dilaton, and a supersymmetric minima is generated for these scalars. However, the K\"ahler directions remain flat and gives the no-scale structure. Quantum corrections, perturbative and non-perturbative, are necessary to understand as they break the no-scale structure and are important for both moduli stabilisation and inflation. The K\"ahler moduli can be fixed by non-perturbative corrections to the superpotential and the most well known ones are the extensively studied KKLT and LVS proposals \cite{Kachru:2003aw, McAllister:2024lnt, Balasubramanian:2005zx, Chauhan:2025rdj}. 

Interestingly, the perturbative corrections give rise to inflation where the inflaton is a K\"ahler modulus. Various constructions involving $\alpha^{\prime}$ and $g_{s}$ corrections remain in literature using the LVS construction \cite{Cicoli:2008gp, Cicoli:2016chb,Brinkmann:2023eph,Bansal:2024uzr}. For example, perturbative string loop corrections were used to study inflation for fibred Calabi-Yau \cite{Cicoli:2008gp}, and whose recent advancements are available in \cite{Cicoli:2016xae, Cicoli:2017axo,Cicoli:2020bao,Burgess:2016owb, Cicoli:2024bxw}. Another interesting version of inflation using blow-up K\"ahler modulus, called K\"ahler Moduli Inflation was studied in \cite{Conlon:2005jm} and subsequent extension of it using string loop corrections, studied in \cite{Bansal:2024uzr}. This was named Loop Blow-up inflation, whose perturbative generalisation we also address in this paper. 

Stabilising moduli's perturbatively was considered in \cite{Burgess:2022nbx} as a general idea depending on logarithmic corrections arising at different orders in $\alpha^{\prime}$ and $g_{s}$. Computation of logarithmic corrections at one loop in $\mathcal{O}(\alpha^{\prime 3})$ \cite{Antoniadis:1997eg, Antoniadis:2018hqy, Antoniadis:2019rkh}, opened up a new avenue to perturbatively stabilise the K\"aler moduli. This was followed by a series of developments in perturbative stabilisation and inflation \cite{Cicoli:2024bwq,Leontaris:2022rzj, Leontaris:2023obe, Bera:2024zsk,Bera:2024ihl,Chakraborty:2025yms, Basiouris:2025yir, Kobayashi:2017zfd, Basiouris:2020jgp, Basiouris:2021sdf,Chakraborty:2025wqn}. 

Perturbatively constructing a dS minima was considered in \cite{Cicoli:2024bwq} in order to resolve the $\eta$-problem in brane-antibrane inflation \cite{Burgess:2001fx, Kachru:2003sx,Majumdar:2002hy,Majumdar:2003da,Villa:2025zmj}. We build on this perturbative stabilisation idea where only a single K\"ahler modulus was considered. In this work, we solely use perturbative corrections to construct a dS vacua and study inflation using only perturbative corrections. Note that the leading order $\alpha^{\prime}$ and $g_{s}$ corrections fix the volume modulus to a dS minima but the other K\"ahler directions remain flat at leading order. These are lifted by moduli-redefinitions and string loop corrections. The article is organised as follows:
\begin{itemize}
    \item In Section \ref{Section 01} we review the single modulus case (i.e. $h^{1,1}=1$), where the volume modulus is stabilised in a de-Sitter minima using only perturbative corrections \cite{Cicoli:2024bwq}.
    \item In Section \ref{section:3}, the construction is extended to include multiple K\"ahler moduli. We illustrate for $h^{1,1}=2$ which, in principle, can be extended to $h^{1,1}>2$. 
    \item Afterwards, in Section \ref{section:4}, we introduce two models of inflation that arise from moduli-redefinitions and string-loop corrections. 
    \item Section \ref{section:5} then discusses the region of parameter space where the two models can operate when EFT constraints and CMB data is taken into account. 
    \item Lastly, in Section \ref{section:6}, we discuss the values of spectral index $n_{s}$ and tensor-to-scalar ratio $r$ of these two models that are compatible with experiments \cite{Planck:2018jri, ACT:2025fju, ACT:2025tim,DESI:2024mwx, Kallosh:2025ijd}. 
\end{itemize}

\section{\texorpdfstring{Perturbative Stabilisation of $h^{1,1} = 1$}{Perturbative Stabilisation of h1,1 = 1}}\label{Section 01}
We work in the four-dimensional Einstein frame. The reduced Planck mass is
\begin{align}
    M_{\mathrm{P}} = \frac{1}{\sqrt{8 \pi G}},
\end{align}
where we have set $M_{\mathrm{P}} = 1$. Furthermore, the string length is defined as $l_{s}=2\pi\sqrt{\alpha^{\prime}}$, and the dimensionless Calabi-Yau volume is defined in the Einstein frame as 
\begin{align}
    \mathcal{V} = \frac{\text{Vol}_{\text{E}}(X_{6})}{l_{s}^{6}} \ .
\end{align}
We expand the K\"ahler form as $J=t^{i}\omega_{i}$, to express the two cycle volume $t_{i}$, the 4-cycle volume $\tau_{i}$, and Complex K\"ahler coordinates $T_{i}$, with the following relations 
\begin{align}
    \mathcal{V}=\frac{1}{6} \kappa_{i j k} t^i t^j t^k, \quad \tau_i=\frac{1}{2} \kappa_{i j k} t^j t^k, \quad T_i=\tau_i+i \theta_i .
\end{align}

One of the central challenges in string phenomenology is the complete stabilisation of all moduli fields. In four-dimensional $\mathcal{N}=1$ supergravity, the dynamics of these moduli are encoded in the F-term scalar potential, which takes the form
\begin{equation}
    V=e^{K}\left(\sum_{A, B}\left(D_A W\right) K^{A \overline{B}}\left(D_{\overline{B}} \overline{W}\right)-3|W|^2\right),
\end{equation}
where $A, B$ run over every chiral multiplet in the theory, $K$ is the Kähler potential, $W$ the superpotential, and $D_{A} W \equiv \partial_A W+\left(\partial_A K\right) W$ the Kähler-covariant derivative. For Type IIB string theory compactified on a Calabi-Yau threefold, the relevant moduli comprise the complex-structure moduli $U^{a}$, the Kähler moduli $T_i$, and the axio-dilaton $S$. The tree-level superpotential is \cite{Gukov:1999ya} 
\begin{align}
    W_{0}=\int_{\text{CY}_{3}}G_{3}\wedge \Omega
\end{align}
where $G_3=F_3-S H_3$ combines the Ramond-Ramond and Neveu-Schwarz three-form fluxes, and $\Omega$ is the holomorphic $(3,0)$. Crucially, $W_{0}$ depends only on $S$ and the $U^a$ through $\Omega$, but is independent of the Kähler moduli $T_{i}$. The tree-level K\"ahler potential is given by
\begin{align}
    K_{0}=-2 \ln \left(\mathcal{V}\right)-\ln \left(S+\overline{S}\right)-\ln \left(-i \int \Omega \wedge \overline{\Omega}\right)
\end{align}
where
\begin{equation}
\mathcal{V}=\frac{1}{6} k_{i j k} t^i t^j t^k
\end{equation}
denotes the Calabi-Yau volume in Einstein frame, with $\tau_{i}=Re (T_{i})$, at tree-level before possible loop induced redefinitions. The axio-dilaton, takes the following form
\begin{equation}
S=e^{-\phi}+i C_{0}=\frac{1}{g_{s}}+i C_{0} \equiv s+i C_{0}
\end{equation}
which allows the use $s$ or $g_{s}$ interchangeably (i.e. $s = \frac{1}{g_{s}}$) to denote the string coupling which we shall do so throughout the discussion of this article. At tree-level, the F-term has a supersymmetric minima for the complex structure moduli and dilaton however the scalar potential has a no-scale structure for the K\"ahler moduli i.e at tree-level the K\"ahler directions remain flat. This can be remedied by either non-perturbative corrections or perturbative corrections. 

We will now highlight the case of a single K\"ahler modulus which can be stabilised by perturbative corrections and can be guaranteed to sit at a dS minima \cite{Cicoli:2024bwq}. Utilising perturbative corrections at $\mathcal{O}(\alpha^{\prime 3})$ and logarithmic loop corrections at $\mathcal{O}(\alpha^{\prime}{}^3,g_{s}^{2})$, a single K\"ahler modulus can be stabilised, and then the uplifting from AdS to dS being achieved via moduli-redefinitions.\footnote{A discussion on perturbative corrections and the scaling of subsequent K\"ahler potential is highlighted in App. \ref{Known Corrections}} The reason this is possible without the breakdown of perturbation theory is that both the corrections at $\mathcal{O}(\alpha^{\prime 3})$ depend on the euler-character of the Calab-Yau and the correction at one-loop $\mathcal{O}(\alpha^{\prime 3})$ has a logarithmic enhancement which covers up for it being subleading in $g_{s}$. For a single K\"ahler modulus, the K\"ahler potential with leading corrections are 
\begin{equation} \label{Kahler pot 1 modulus}
    K \simeq -3 \ln{\left[\tau-\alpha \ln{\tau} + \frac{\xi s^{3/2}}{3\sqrt{\tau}} - \frac{D \ln{\tau} }{s^{1/2}\sqrt{\tau}}\right]},
\end{equation}
where, $\alpha$ is the moduli-redefinition coefficient corresponding to the volume modulus, $\xi$ the BBHL $\mathcal{O}(\alpha^{\prime 3})$ correction and $D$ the logarithmic correction at $\mathcal{O}(\alpha^{\prime}{}^3,g_{s}^{2})$. The details of these corrections are fleshed out in the App. \ref{Known Corrections}. The perturbative string loop corrections \cite{Berg:2005ja} were checked to be subleading in \cite{Cicoli:2024bwq}. Plugging the above K\"ahler potential with the tree-level superpotential into the F-term scalar potential results in 
\begin{equation}\label{Scalar pot 1 modulus}
    \frac{V}{3 |W_0|^2} = \frac{\alpha}{\tau^4}+\frac{s^{2}\xi+\mathcal{O}(1/s^2)-3D\ln{\tau}}{4 s^{1/2}\tau^{9/2}}+ \mathcal{O}\left({\frac{1}{\tau^5}}\right),
\end{equation}
this potential then admitting a dS minimum at 
\begin{equation}
\tau_{\min } \sim e^{\frac{c}{g_{s}^2}} \quad \text{where} \quad c \equiv \frac{\xi}{D}=\frac{\zeta(3)}{3 \zeta(2) T_{7}} \ ,
\end{equation}

provided $\alpha$ is not too large, as is expected as it is a one-loop $\beta$ function, and us stabilising perturbatively the K\"ahler moduli (saxions). However, axionic directions remain flat, since they cannot be fixed via perturbative corrections. They require non-perturbative corrections to be stabilised, and is found to be subleading \cite{Cicoli:2024yqh, Chakraborty:2025wqn}. Note that such corrections (like poly-instanton corrections) are subleading in our setup, and do no actively effect the inflationary dynamics but can lead to interesting phenomenology as explored in the cited papers above. This specific perturbative stabilisation methodology in \cite{Cicoli:2024bwq} helped overcome the $\eta$-problem in brane-antibrane inflation. However, a discussion of stabilisation of multiple K\"ahler modulus was beyond the scope of the analysis of the article. We expand upon that in the following sections.

\section{Stabilisation of multiple K\"ahler moduli}\label{section:3}
We generalise the perturbative stabilisation idea introduced in the last section, to include a blow-up K\"ahler modulus. Considering a Calabi-Yau with the following volume form 
\begin{equation}\label{Volume CY general}
    \mathcal{V}= \frac{1}{\gamma} \left( \tau_b^{3/2} - \lambda_s \tau_s^{3/2}\right).
\end{equation}
Here $\tau_{b}$ denotes the large cycle volume, $\tau_{s}$ denotes the small cycle volume, with $\tau_{b} >> \tau_{s}$. Furthermore, $\gamma$ represents an overall normalisation constant, and $\lambda_{s}$ represents the ratios of triple intersection numbers. Explicitly, 
\begin{align}
\begin{array}{ll}
\frac{1}{\gamma}=\frac{\sqrt{2}}{3 \sqrt{\kappa_{b b b}}}, & \frac{\lambda_s}{\gamma}=\frac{\sqrt{2}}{3 \sqrt{\kappa_{s s s}}}, \\
\gamma=3 \sqrt{\frac{\kappa_{b b b}}{2}}, & \lambda_s=\sqrt{\frac{\kappa_{b b b}}{\kappa_{s s s}}} ,
\end{array}
\end{align}
where $k_{bbb}$ and $k_{sss}$ are said triple intersection numbers of divisors on the Calabi–Yau.

\begin{figure}[h]
    \hspace{0.2in}
    \includegraphics[width=1\linewidth]{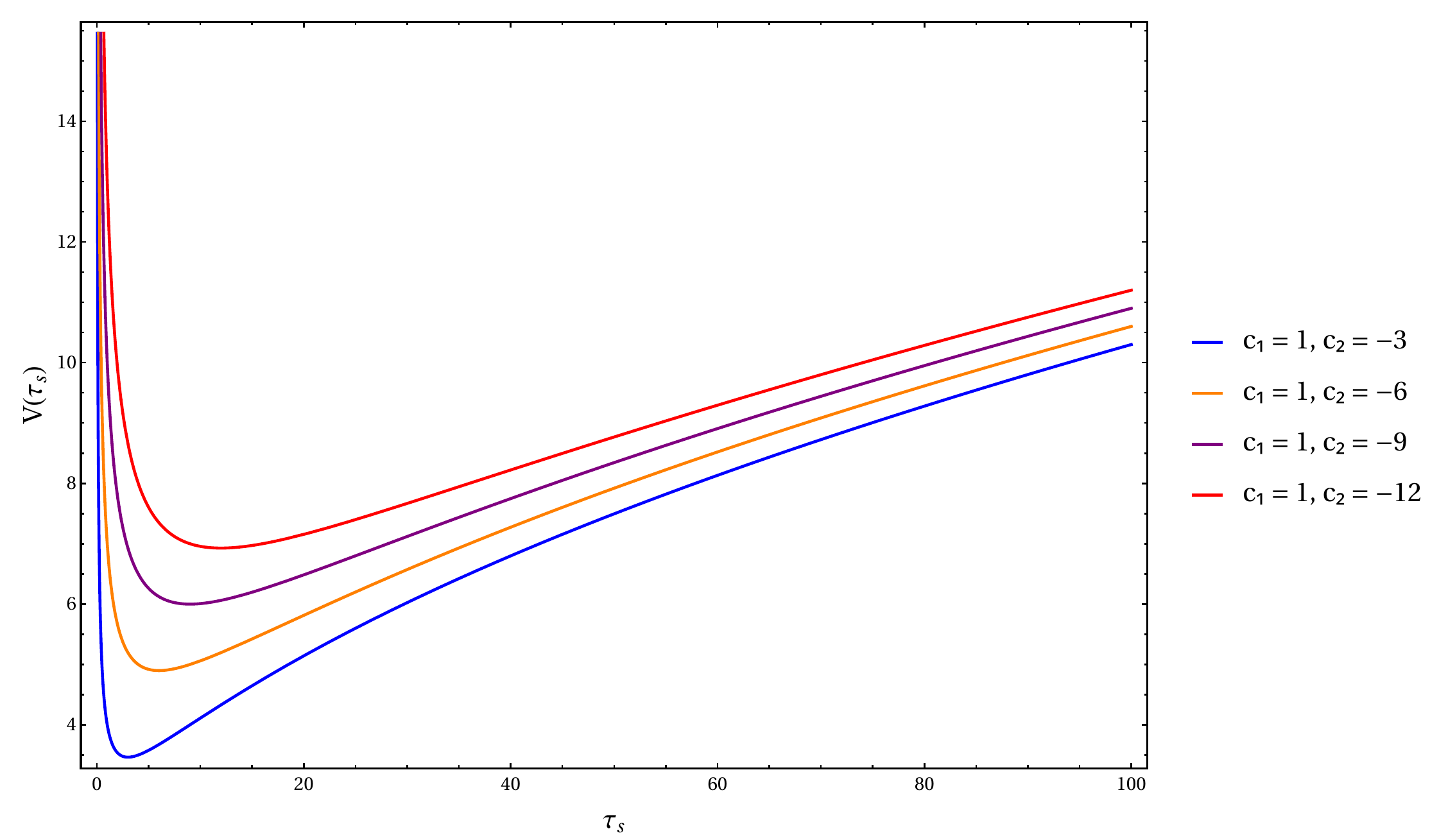}
    \caption{In the above plot, we considered the potential $V(\tau_{s})=\left(
          c_{1}\,\sqrt{\tau_{s}}
          - \frac{c_{2}}{\,\sqrt{\tau_{s}}}\right)$ and also kept $c_{1}$ fixed. In the above plot, which is just for illustration purposes, it is evident as $c_2$ takes on larger values, the minima of the potential moves to a larger value, which is evident from \eqref{tau_min}. The minima of the red curve is at $\tau_s=12$, and as we subsequently decrease $c_2$, the minima move to $\tau_s=3$ for the blue curve. If either one of the cases in the potential is to serve as an inflaton candidate, one can tune the other coefficient as to take a value of $\tau_{s}$ satisfying inflationary and K\"ahler cone constraints.}
    \label{fig:tau s plot}
\end{figure}

To aid in the computation, we consider the division of the K\"ahler potential \cite{Cicoli:2007xp} as 
\begin{equation}\label{Kahler pot 0}
    K_{0}=-2 \ln \left(\frac{\left(\tau _b-\alpha  \ln \left(\tau _b\right)\right){}^{3/2}-\lambda_s \left(\tau _s+\beta\left(\ln\tau _b\right)\right)^{3/2}}{\gamma }-\frac{1}{2} \xi  s^{3/2} + D s^{-1/2} \ln{\tau_b} \right). 
\end{equation}
and 
\begin{equation}\label{Kahler pot 1}
    K_{loop}=\frac{A_b \sqrt{\tau _b}}{s \tau _b^{3/2}}+\frac{A_s \sqrt{\tau _s}}{s \tau _b^{3/2}}+\frac{B_b}{s \tau _b^{2}}+\frac{B_s}{s \sqrt{\tau _s} \tau _b^{3/2}} \ .
\end{equation}
In the K\"ahler potential equation  \eqref{Kahler pot 0}, we have considered moduli redefinitions \cite{Conlon:2009kt, Conlon:2010ji} of both the K\"ahler moduli (which we denote by $\alpha$ and $\beta$ \footnote{Logarithmic moduli redefinitions were observed for the small four cycle moduli $\tau_s$ in \cite{Conlon:2010ji}, and for the large four cycle moduli in \cite{Klaewer:2020lfg}. However, since this is not explicitly derived in the paper for both moduli, we take this as more of a phenomenological motivation.}) along with the leading order $\alpha^{\prime 3}$ correction \cite{Becker:2002nn} and one-loop $\alpha^{\prime 3}$ logarithmic correction \cite{Antoniadis:2018hqy,Antoniadis:2019rkh,Leontaris:2022rzj}. This would stabilise the big-cycle modulus analogous to the single modulus case of Sec. \ref{Section 01}. As discussed in App. \ref{Known Corrections}, $A$ denote the coefficients of string loop corrections due to KK-modes and $B$ denote the coefficient of string loop corrections due to winding both and both of them are functions of the complex-structure moduli and dilaton. Due to the extended no scale \cite{Cicoli:2007xp}, string loop corrections appear subleading and, in our notation, it is given by equation  \eqref{Kahler pot 1}. Following the footsteps of the analysis \cite{Cicoli:2007xp} and up to second order in loop corrections, we get the following,
\[
\left\{
\begin{aligned}
\delta V_1 &= \left( 2 K_0^{ij} K_i^0 \delta K_j - K_0^{im} \delta K_{ml} K_0^{lj} K_i^0 K_j^0 \right) \frac{|W_{0}|^2}{\mathcal{V}^2} \\
\delta V_2 &= \left( K_0^{ij} \delta K_i \delta K_j - 2 K_0^{im} \delta K_{ml} K_0^{lj} K_i^0 \delta K_j \right. \\
&\quad \left. + K_0^{im} \delta K_{mp} K_0^{pq} \delta K_{ql} K_0^{lj} K_i^0 K_j^0 \right) \frac{|W_{0}|^2}{\mathcal{V}^2}.
\end{aligned}
\right.
\]
Using the above equations, we get the following leading-order contribution
\begin{equation}
\frac{V_{0}}{3\,|W_{0}|^2}\simeq\frac{\alpha  \gamma ^2}{\tau_b^4}-\frac{\gamma ^3 \xi  s^{3/2}}{4 \tau_b^{9/2}}  +\frac{\gamma ^3 s^{-1/2}D \ln{\mathcal{V}}}{4 \tau_b^{9/2}}-\frac{2\gamma^3 Ds^{-1/2} }{\tau^{9/2}_{b}}\label{bigg}
\end{equation}
equation  \eqref{bigg} leads to the stabilisation of the $\tau_{b}$ modulus that is analogous to the method highlighted in the previous section, courtesy of equation equation  \eqref{Scalar pot 1 modulus}.\footnote{One might ask the question as to how equation  \eqref{Scalar pot 1 modulus} is analogous to equation  \eqref{bigg}. Note that, the term proportional to $D$ without logarithm is one-loop suppressed compared to the BBHL term. It is shown in the appendix of \cite{Cicoli:2024bwq} that such terms do not ruin the minima construction.} The F-term scalar potential up to second order in loop corrections is given by 
\begin{equation}
\begin{aligned}
\frac{V_{\text{sub}}}{3\,|W_{0}|^2}
&\simeq
\frac{\gamma^2\bigl(-4A_sB_s\,\tau_s + A_s^2\,\tau_s^2 + 4B_s^2\bigr)}
     {6\,s^2\,\tau_b^{9/2}\,\tau_s^{5/2}\lambda_s}
+ \frac{\beta\,\gamma^2\bigl(A_s\,\tau_s - A_s\,\tau_s\ln\tau_b + 2B_s\ln\tau_b - 2B_s\bigr)}
       {s\,\tau_b^{9/2}\,\tau_s^{3/2}\lambda_s} \nn \\
&\quad
+ \frac{\lambda_s\beta\,\gamma^2\,\tau_s^{1/2}}{\tau_b^{9/2}}\,.
\end{aligned}
\end{equation}
Note that at order $\mathcal{O}\left(\frac{1}{\tau_b^{9/2}}\right)$, there is an expansion in $\frac{1}{\tau_s^{1/2}}$ going on. This lets us stabilise $\tau_s$ using leading order corrections. The potential becomes
 \begin{equation} \label{ts pot}
    \frac{V(\tau_s)}{ \gamma^2 |W_{0}|^2} \simeq \frac{3\lambda_s\beta\sqrt{\tau_s}}{\tau_b^{9/2}} + \frac{\left( - 12 \lambda_s s B_s 
    + A_s^2 \right)} { 6s^{2}\lambda_s\sqrt{\tau_s} \tau_b^{9/2}} \ .
\end{equation}
It is easy to see that the above potential contains a minima for $\tau_{s}$. The full potential with $\tau_{b}$ stabilised can be written as
\begin{equation}
V
= V_0
+ \frac{ \gamma^2|W_0|^2}{\tau_b^{9/2}}\left(
          3\lambda_s\beta\sqrt{\tau_s}
          + \frac{\left(-12s\,B_s \lambda_s + A_s^2\right)}{6s^2\lambda_s\sqrt{\tau_s}}
        \right)
        \,
\end{equation}
where $V_{0}$ denotes the minima of $\tau_{b}$ where 
\begin{equation}\label{minima of tau big during inflation}
V_{0}=V\left(\tau_{b ;\ min}\right) \simeq \frac{3\gamma^3 W_{0}^2\xi s^{3 / 2}}{4\tau_{b ;\ min }^{9 / 2}}\equiv\frac{\gamma^2 W_{0}^2 V_{min}}{\tau_{b;\ min }^{9 / 2}} \ .
\end{equation}
We present $V$ in a more convenient form to aid in the inflationary analysis. 
\begin{equation}\label{convenient form}
V\simeq V_{0} \left(1 + \frac{1}{V_{min}} \left(c_{1}\,\sqrt{\tau_{s}}-\frac{c_{2}}{\,\sqrt{\tau_{s}}}\right)\right),
\end{equation}
where, $c_{1}=3\lambda_s\beta$ and $\displaystyle c_{2}=\frac{c_{loop}}{\lambda_s}$. The potential for $\tau_{s}$ reaches a minima when 
\begin{equation}\label{tau_min}
    \tau_{s} = -\frac{c_{2}}{c_{1}} 
\end{equation}
with either of the loop coefficients being negative. In Fig. \ref{fig:tau s plot}, we plot $V(\tau_s)$ vs $\tau_s$. 

The procedures adopted in the above discussions can in principle be extended to models containing multiple K\"ahler modulus. In that case, one way to stabilise multiple blow-up modulus would be to incorporate a moduli redefinition of each of the blow-up modulus. As evident from our analysis, this would stabilise all the blow-up modulus. In the next section, we address the inflationary models arising from the above analysis. 
\section{The Inflationary Models}\label{section:4}
The potential \eqref{convenient form} gives rise to two distinct classes of inflationary models with their own distinct parameter space, thus leading to non-identical inflationary predictions.\footnote{Refer to \cite{Cicoli:2011zz}, so as to achieve a concrete review of inflation using string moduli.} The two cases are two limits of the complete potential in \eqref{convenient form} rather than unrelated potentials. The first regime applies when: 
\begin{equation}
    \left|c_2\right| \ll\left|c_1\right| \tau_s.
\end{equation}
This entails the scenario where the term related to the moduli redefinitions of $\tau_{s}$, truncates over second term in the full potential generated by string loop corrections.

In similar fashion, we find the second regime to apply when 
\begin{equation}
    \left|c_1\right| \tau_s \ll\left|c_2\right|,
\end{equation}
here the loop-corrected term domanates over the moduli redefinition of $\tau_{s}$. Thus generating two distinct cases of inflationary potentials with their unique predictions. After inflation, $\tau_s$ settles in it's minima \eqref{tau_min}.

The first class of models arise due to the moduli redefinition of the small blow up cycle $\tau_s$ and it gives rise to a potential corresponding to the first term in equation  \eqref{convenient form}. The second class of the models, which is basically the second term of equation  \eqref{convenient form}, are analogous to \cite{Bansal:2024uzr} as they arise from string loop corrections. The canonically normalised field inflaton field is \cite{Conlon:2005jm,Bansal:2024uzr}
\begin{align}
    \tau_{s}=\left(\frac{3 \mathcal{V}}{4 \lambda_{s}}\right)^{2/3}\phi^{4/3}.
\end{align} 
\subsection{Case 1}
The potential, considering only the first term in equation  \eqref{convenient form} is rewritten as 
\begin{align} \label{case 1 pot} 
V=V_0\left(1+\frac{1}{V_{\text {min }}}\left(C_1 \phi^{2 / 3}\right)\right), 
\end{align}
where, $C_1\equiv c_1\left(\frac{3 \tau_b^{3/2}}{4 \gamma \lambda_{s}}\right)^{1 / 3}$. This gives rise to the following slow-roll parameters
\begin{align}
    \epsilon &\simeq \frac{2}{9} \frac{1}{V_{min}^2} C_1^2 \phi^{-2/3} \nn \\
    \eta &\simeq - \frac{2}{9V_{min}} C_1\phi^{-4/3} .
\end{align}
The number of e-foldings
\begin{align}
    N=\int^{\phi_{*}}_{\phi_{end}} d\phi \frac{V(\phi)}{V^{\prime}(\phi)}\ ,
\end{align}
give us an expression for horizon exit $\phi_{*}$, 
\begin{equation}\label{phi star 1}
    \phi_{*} = \left(\frac{8 N}{9 V_{min}}C_1\right)^{3/4}.
\end{equation}
One can easily check that this supports slow-roll inflation. It is important to investigate if the potential in consideration can produce enough density perturbations with tensor-to-scalar ratio and spectral tilt that coincide with current observational bounds. We will address this in section \ref{section:5}. We will now focus on the second term in Eq \eqref{tau_min} and call it case 2. 
\subsection{Case 2}
Considering the second case, the potential with canonically normalised inflaton takes the form 
\begin{align} \label{case 2 pot} 
V=V_0\left(1-\frac{1}{V_{\text {min }}}C_2 \phi^{-2 / 3}\right),
\end{align}
where $C_2=c_2\left(\frac{3 \tau_b^{3/2}}{4 \gamma \lambda_{s}}\right)^{-1 / 3}$. This leads to the following slow-roll parameters
\begin{align}
    \epsilon &\simeq \frac{2}{9} \frac{1}{V_{min}^2} C_2^2  \phi^{-10/3} \nn \\
    \eta &\simeq - \frac{10}{9V_{min}} C_2 \phi^{-8/3} \ . 
\end{align}
In similar fashion, the number of e-foldings $N$ give us an expression for horizon exit $\phi_*$, 
\begin{equation}\label{phi star 2}
    \phi_* = \left(\frac{16 N}{9 V_{min}} C_2 \right)^{3/8}.
\end{equation}
Again, this also supports slow-roll inflation, and we will analyse if this model, as well produce enough e-foldings that will describe inflation within current bounds.

In the next section, we will work with both case $1$ and case $2$ by imposing constraints given by observational bounds and show how we can obtain allowed parameters to produce inflation from both models. 
\section{Constraints}\label{section:5}

In this section, we temporarily restore the reduced Planck mass $M_P$ in order to make the mass dimensions explicit; elsewhere we continue to use the convention $M_P=1$. There are various kinds of constraints which we will incorporate in our setup arising from string theory and observations. The first constraint is the inflationary constraints that we encounter in order to match the amplitude of density perturbations. This would fix one of our parameters in the analysis as we will see. Other kinds of constraints include string couplings within a tolerable range, a not too small volume and the supersymmetry breaking scale well below the KK scale. Finally, another stringent constraint comes from having the whole inflationary dynamics within the K\"ahler cone. For this, we need to specify $\gamma$ and $\lambda_{s}$. We choose the following form of the volume
\begin{equation}
\mathcal{V}=\frac{1}{9} \sqrt{\frac{2}{3}}\left(\tau_b^{3 / 2}-\sqrt{3} \tau_s^{3 / 2}\right).
\end{equation}
Thus, our $\gamma=9 \sqrt{\frac{3}{2}}$ and $\lambda_{s}=\sqrt{3}$ in this case. The Kähler cone conditions are basically
\begin{equation}
    t_{b}+t_{s}>0, \quad t_s<0.
\end{equation}
Using canonical normalisation, we can arrive at the expression
\begin{equation}
\tau_s \simeq\left(\frac{1}{18 \sqrt{2}}\right)^{2 / 3} \, \phi^{4 / 3}\tau_b.
\end{equation}
Since the K\"ahler cone constraints are in terms of 2-cycles, we can rewrite as 
\begin{equation}
\frac{\left|t_s\right|}{t_b}=\left(\frac{1}{2 \sqrt{6}}\right)^{1 / 3} \phi_*^{2/3} \simeq 0.589 \hspace{1mm} \phi_*^{2/3}\ .
\end{equation}
As inferred from the above expression, we need to satisfy $\phi_* \lesssim 1$ to be inside the K\"ahler cone. In summary, we impose the following constraints in both the cases whilst studying their parameter space. 
\begin{enumerate} \label{List of constraints}
    \item We \textit{impose} a conservative constraint on the string coupling $s>5$ which is $0<g_{s}<0.25$. This ensures that we are at the weak coupling regime and as a positive consequence, since our volume modulus scales as $e^{1/g_s^2}$, we also avoid small volumes by being in this regime. 
    \item The number of e-foldings is \textit{enforced} to be within $40<N<60$. 
    \item The spectrum of density perturbations \cite{Planck:2018jri} is defined as
    \begin{equation}
\Delta_s^2=\tilde{A}_s\left(\frac{k}{k_*}\right)^{n_s-1}.
\end{equation} 
Where the amplitude $\tilde{A_{s}}$ is defined as follows
\begin{align}
    \tilde{A_s}=\left.\frac{V_{\mathrm{eff}}}{24 \pi^2 M_{P}^4 \ \epsilon_V}\right|_*,
\end{align}
where, as measured by PLANCK \cite{Planck:2018jri} is
\begin{equation}\label{A hat}
\hat{A}_s \equiv 12 \pi^2 \tilde{A}_s \simeq 2.5 \times 10^{-7}. 
\end{equation}
 We can relate this to the potential and the amplitude of density perturbations \cite{Planck:2018jri} at horizon exit can be written as
\begin{equation} \label{constraints density pert}
\left.\frac{V^3}{V_\phi^2}\right|_{\phi=\phi_*}=\hat{A}_s \equiv 12 \pi^2 \tilde{A}_s \simeq 2.5 \times 10^{-7}
\end{equation}

whereby our parameters are fixed in the subsequent subsection.
    \item Kaluza-Klein scale larger than the SUSY breaking scale is also imposed. In $\mathcal{N}=1$ supergravity the total SUSY-breaking F-term sets the gravitino mass i.e. $m_{3 / 2}=e^{K / 2}|W_{0}| \sim \frac{W_{0}}{\mathcal{V}} M_{\mathrm{P}}$. The KK-mass is given by $m_{K K} \sim \frac{1}{R} \sim \frac{M_{\mathrm{P}}}{\mathcal{V}^{2 / 3}}$. In particular, as pointed out in~\cite{deAlwis:2012vp,Cicoli:2013swa}, a consistent expansion in superspace requires that the KK scale and the gravitino mass are related as
\begin{equation}\label{eq:gravitino}
\frac{m_{3/2}}{m_{KK}} \sim \frac{W_{0}\sqrt{g_s}}{\tau_b^{1/2}}\ll 1. 
\end{equation}
In our analysis we \textit{impose} the numerical constraint $\frac{W_{0}\sqrt{g_s}}{\tau_b^{1/2}}<\frac{1}{4}$. 
    \item The minima of our potential for $\tau_b$ is at equation  \eqref{minima of tau big during inflation} and is imposed to be fixed during inflation. The minima of this potential depend on the parameter $\xi$ which descends from the $\alpha^{\prime}{}^{3}$ and $(Riem)^4$ correction in 10 dimensions as highlighted in App. \ref{Known Corrections}. This correction depends on the Euler characteristic of the Calabi-Yau as 
    \begin{equation}
    \xi =\frac{\zeta(3) \chi(X)}{(2 \pi)^3 } \ .
\end{equation}
    We can see that the euler character of the CY peaks when $|\xi| \in (0.1,1.5)$ \cite{Kreuzer:2000xy}. In all the subsequent analysis, we operate in the region where most Calabi-Yaus have been found. 
    \item Finally, as discussed, the K\"ahler cone conditions require $\phi_{*} \lesssim 1$, to which we further \textit{impose} $\phi_*<1$ in order to be well inside the K\"ahler cone. 
\end{enumerate}
\subsection{Case 1}
\begin{figure}[h]
    \hspace{0.6in}
    \includegraphics[width=0.95\linewidth]{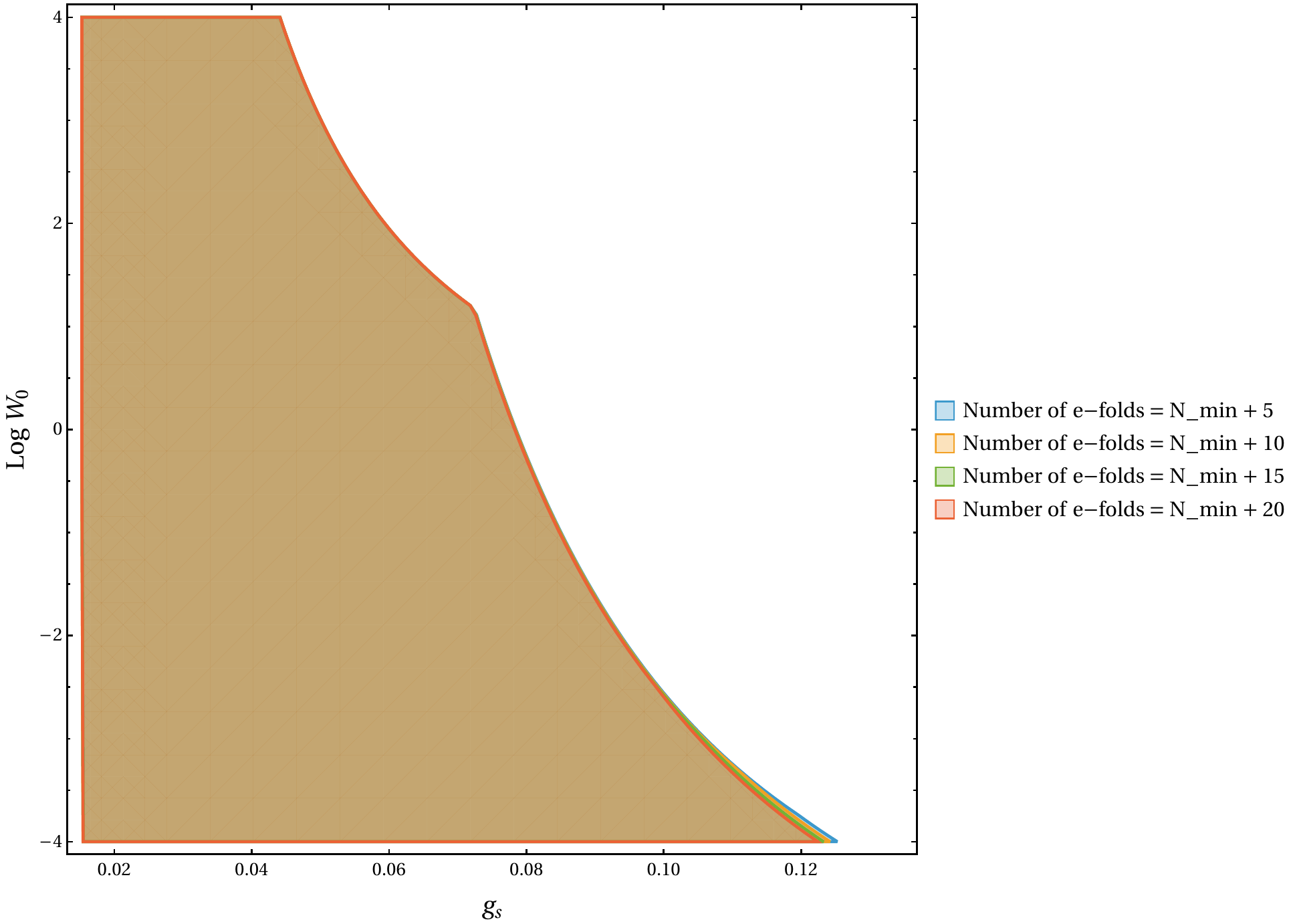}
    \caption{With the constraints discussed in \ref{List of constraints}, the region of parameter space for case 1 is presented with all constraints satisfied. The vertical axis uses a 10-based logarithm. This region plot corresponds to $\xi=0.5$. In principle, one can get such a region for each value of $\xi$ as allowed by \ref{List of constraints}.}
    \label{fig:Case 1 region plot}
\end{figure}

For our first case, we match density perturbations courtesy of equation  \eqref{constraints density pert} and rewrite $C_{1}$ as 
\begin{equation}\label{c1 density pert}
    C_{1} = \left(\frac{3W_{0}^2 \gamma^2 \sqrt{8N} V_{min}^{5/2}}{4\tau_{b}^{9/2}\hat{A}_{s}}  \right)^{2/3},
\end{equation}
and impose equation  \eqref{A hat} in the numerical analysis. The K\"ahler cone constraint $\phi_{*} <1$ in terms of our parameters is obtained by substituting $C_{1}$ in the expression for $\phi_{*}$
\begin{align}
  \left(\frac{8 N}{9 V_{min}}\left(\frac{3W_{0}^2 \gamma^2 \sqrt{8N} V_{min}^{5/2}}{4\tau_{b}^{9/2}\hat{A}_s}  \right)^{2/3}\right)^{3/4} &< 1. 
\end{align}
We defined $C_{1}\equiv c_{1}\left(\frac{3 \tau_b^{3/2}}{4 \gamma \lambda_{s}}\right)^{1 / 3}$, where $c_{1} = 3 \lambda_{s} \beta$. Since $\beta$ is a 1-loop $\beta$-function coefficient of a $SU(N)$ gauge theory (along with some order one-factors), we impose the constraint for $\beta$ as $0<\beta<1$. Moreover, we \textit{impose} the constraint such that inflationary dynamics at larger values of $\phi$ (or $\tau_s$) do not ruin the minima of the volume modulus $V_0$ \footnote{We thank Michele Cicoli for pointing this out.}
\begin{equation}\label{Minima security}
    \frac{V_{c_1}}{V_0}<10^{-2} \quad \text{with} \quad V_{c_1} \equiv \frac{3\gamma^2|W_{0}|^2 c_1}{\tau_b^{9/2}}
          \sqrt{\tau_s} \ .
\end{equation}
Imposing the constraints listed in \ref{List of constraints} and subsequent constraints detailed in this section, we find a promising region of parameter space in Fig. \ref{fig:Case 1 region plot} where all constraints are satisfied. 
\subsection{Case 2}
\begin{figure}[h]
    \hspace{0.5in}
    \includegraphics[width=1\linewidth]{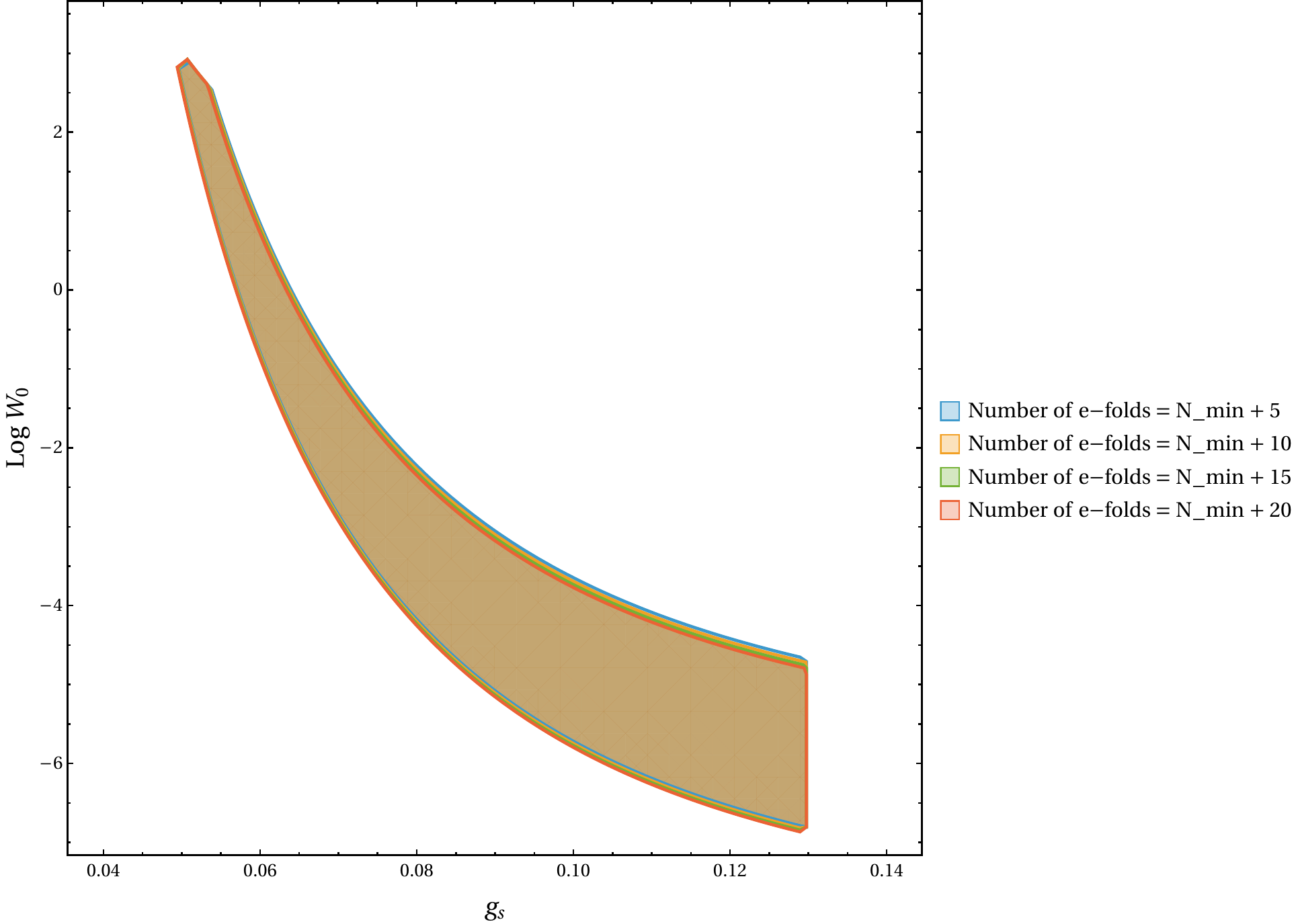}
    \caption{With the constraints \ref{List of constraints}, region of parameter space for case 2 with all constraints satisfied. This region plot corresponds to $\xi=0.5$. The vertical axis uses a 10-based logarithm. In principle, one can get such a region for each value of $\xi$ as allowed by \ref{List of constraints}.}
    \label{fig:Case 2 region plot}
\end{figure}

Similar to the first case, we can use density perturbations equation  \eqref{constraints density pert}, to fix $C_2$ as
\begin{equation}\label{c2 density pert}
      C_2 = \left(\frac{9W_{0}^{2} \gamma^2 V_{min}^{7/4}}{4\tau_{b}^{9/2}\hat{A}_s} \left(\frac{16N}{9 V_{min}}\right)^{5/4} \right)^{4/3}.
\end{equation}
Here, the K\"ahler cone constraint $\phi_* <1$ gives us the following condition 
\begin{equation}
   \left(\frac{16 N}{9 V_{min}}  \left(\frac{9W_{0}^{2} \gamma^2 V_{min}^{7/4}}{4\tau_{b}^{9/2}\hat{A}_s} \left(\frac{16N}{9 V_{min}}\right)^{5/4} \right)^{4/3} \right)^{3/8} <1,
\end{equation}
where we have substituted $C_2$ in the expression for $\phi_*$. Lastly, we defined $C_2=c_2\left(\frac{3 \tau_b^{3/2}}{4 \gamma \lambda_{s}}\right)^{-1 / 3}$ where $c_2$ is the string loop correction coefficient. We impose the constraint on $c_2$ as  $10^{-6}<c_2<10$. This would ensure that string loop corrections do not become too large and spoil the minima for $\tau_b$. We do not impose the constraint equation  \eqref{Minima security} as this would be redundant due to the Case 2 potential being evidently subleading for reasonable values of $c_2$. Keeping the above constraints in mind, we plot the region of parameter space in Fig. \ref{fig:Case 2 region plot} where inflation, matching cosmological data, along with the EFT constraints highlighted in the preceding discussions, can occur. 

\section{Inflationary Analysis}\label{section:6}
As evident from the analysis in the preceding section, a region of parameter space for both cases imply that both models are capable of producing required density perturbations in conjunction with the satisfied EFT and consistency constraints. However, matching the spectral index $n_{s}$ and satisfying the bound of the tensor-to-scalar ratio $r$ can be a challenging task. The bound on the tensor-to-scalar ratio $r$ is \cite{Tristram:2021tvh,Planck:2018jri}
\begin{equation} \label{r bound}
r<0.032 \text { at } 98 \% \mathrm{CL}.
\end{equation}
The ACT collaboration \cite{ACT:2025fju, ACT:2025tim} alongside PLANCK \cite{Planck:2018jri} and DESI \cite{DESI:2024mwx, DESI:2024uvr} significantly constraint the spectral index $n_{s}$ as summarised succinctly in \cite{Kallosh:2025ijd}
\begin{equation}\label{ns bound}
n_{s}=0.9743 \pm 0.0034. 
\end{equation}
\subsection{Case 1}
\begin{figure}[h]
    \hspace{0.5in}
    \includegraphics[width=1\linewidth]{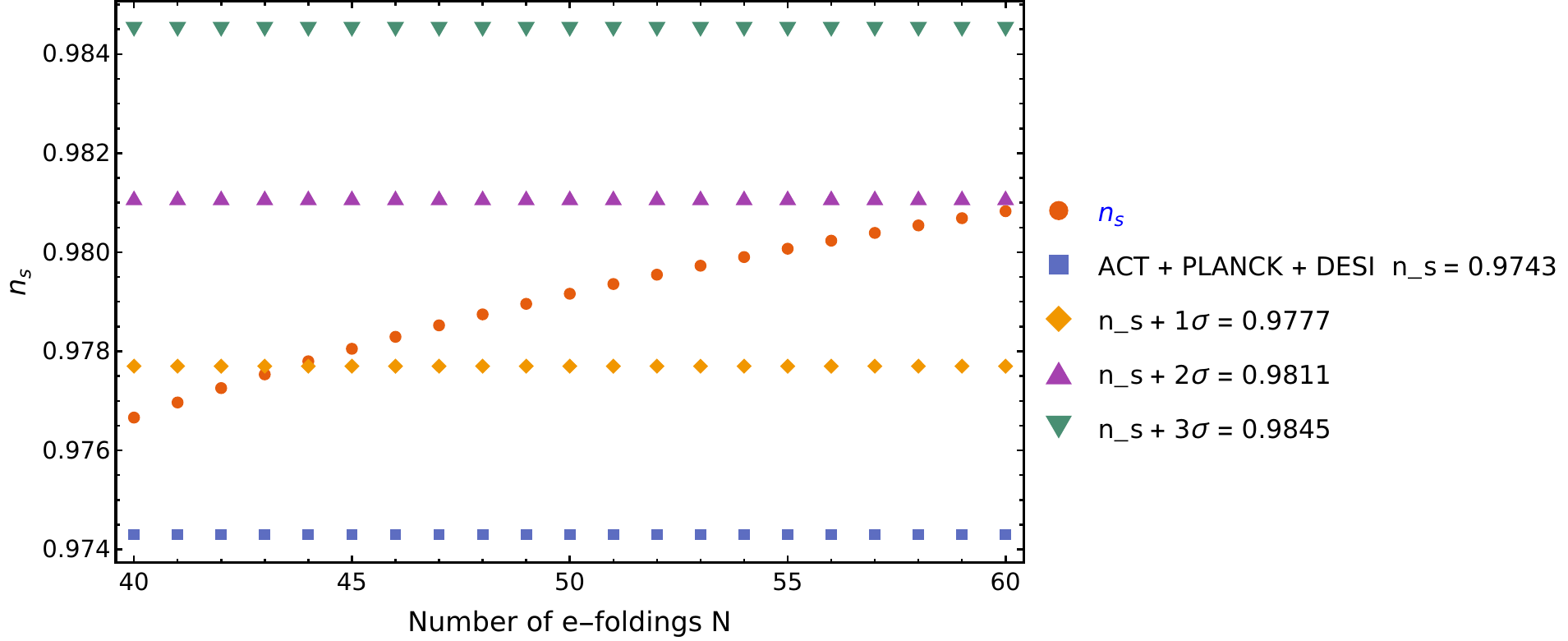}
    \caption{Picking one allowed set of values from Fig. \ref{fig:Case 1 region plot} which is $\displaystyle g_{s}=\frac{1}{14.13}$, $W_{0}=7$ and $\xi=0.5$, we made a $n_{s}$ vs $N$ plot. One can see that the value of $n_s$ approaches close to $n_s+2\sigma$ around $60$ e-folding.}
    \label{fig: ns and N plot}
\end{figure}

\begin{figure}[h]
    \hspace{0.4in}
    \includegraphics[width=1\linewidth]{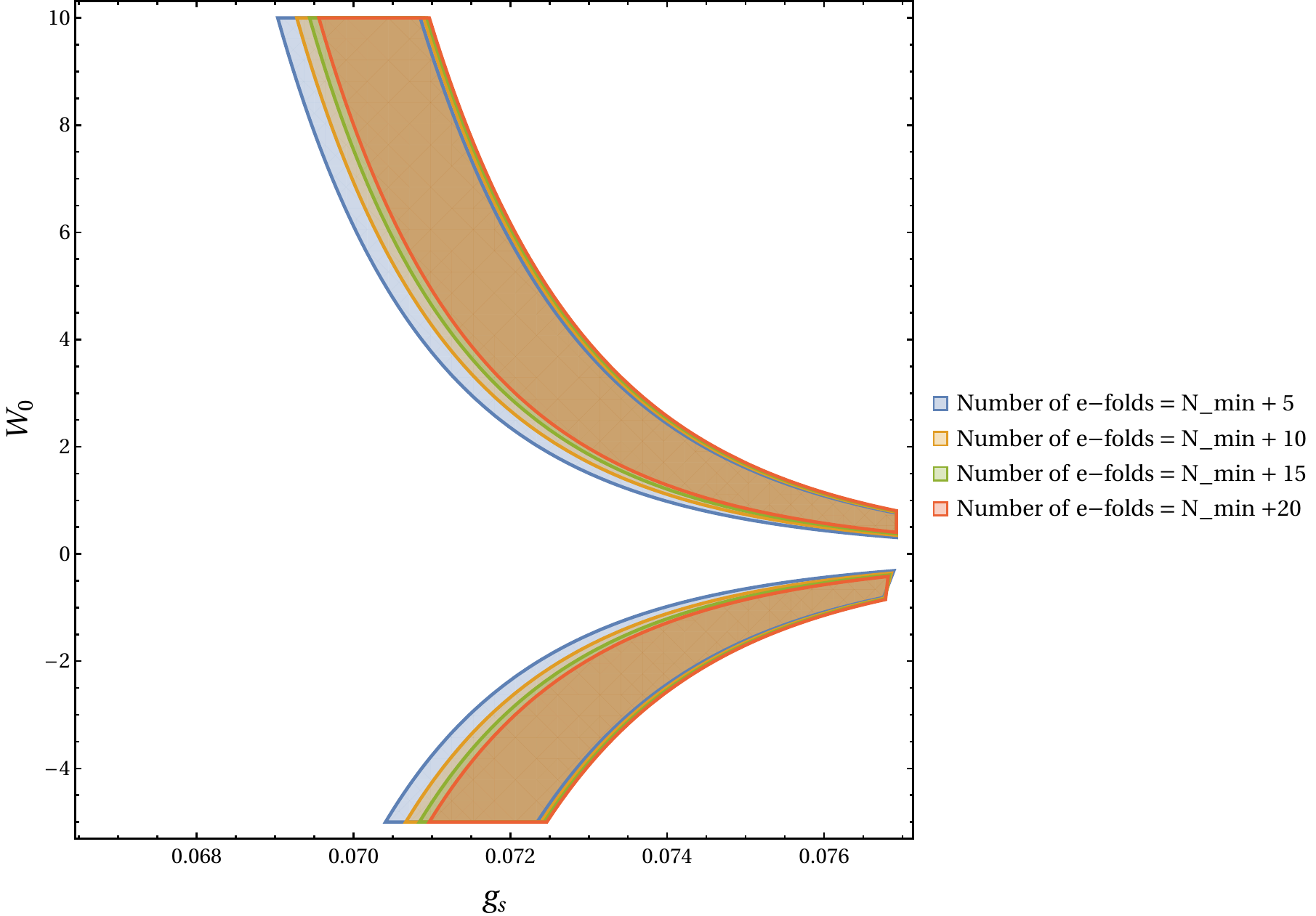}
    \caption{A $W_0$ vs $g_s$ plot ensuring $r<0.032$ and $n_{s}$ within 3$\sigma$ of equation  \eqref{ns bound}.}
    \label{fig: ns and r region plot 1}
\end{figure}

The tensor-to-scalar ratio $r$  and spectral tilt $n_{s}$ is defined by
\begin{align}
    r&=16 \epsilon \nn \\
    n_s&=1+2\eta_*-6\epsilon_*\ .
\end{align}
We can now use the expression for $\phi$ at horizon exit equation  \eqref{phi star 1} and $C_{1}$ in terms of density perturbations equation  \eqref{c1 density pert} to arrive at
\begin{align}
    r&=\frac{8\gamma^2 W_{0}^2 V_{min}}{3 \hat{A}_s \tau_{b}^{9/2}} \nn \\
    n_s&\simeq 1- \frac{1}{2N} -  \frac{3\gamma^2 W_{0}^2V_{min}}{\hat{A}_s \tau_{b}^{9/2}} \ .
\end{align}
Let's take a representative case in the allowed region. Considering the region in Fig. \ref{fig:Case 1 region plot}, we can plot an $n_{s}$ vs $N$ plot Fig. \ref{fig: ns and N plot}. 
In the above case $g_{s}=\frac{1}{14.13}$, $W_{0}=7$ and $\xi=0.5$. This leads us to the following tensor-to-scalar ratio and value of $n_s$ at $54$ e-folds
\begin{equation}
    r \simeq 0.028 \,\,\,\text{and}\,\,\, n_s\simeq 0.979 .
\end{equation}
Another candidate case for $\displaystyle g_{s}=\frac{1}{14.13}$, $\xi=0.5$ and $W_{0}=4$ for which we get
\begin{equation}
    r \simeq 0.0094\,\,\,\text{and}\,\,\, n_{s} \simeq 0.986.
\end{equation}
It is evident that these can give rise series of predictions allowed by the constraints highlighted in \ref{List of constraints} and in Fig. \ref{fig:Case 1 region plot}. However, not all regions in Fig. \ref{fig:Case 1 region plot} satisfy the constraints of $n_{s}$ (equation  \eqref{ns bound}) and $r$ (equation  \eqref{r bound}). With the requirements of $n_{s}$ within $3\sigma$ of equation  \eqref{ns bound} and $r<0.032$, we plot a far more restricted region in Fig. \ref{fig: ns and r region plot 1}.
\subsection{Case 2}
Similarly, we can find the tensor-to-scalar ratio $r$  and spectral tilt $n_{s}$ by using equation  \eqref{phi star 2} and equation  \eqref{c2 density pert}
\begin{align}
    r&=\frac{8\gamma^{2} W_{0}^{2} V_{min}}{\hat{A}_s \tau_{b}^{9/2}} \nn \\
    n_s&=1+2\eta_*-6\epsilon_* \simeq 1- \frac{5}{4N} -  \frac{3\gamma^2 W_{0}^2 V_{min}}{\hat{A}_s \tau_{b}^{9/2}} \ .
\end{align}
In this scenario, the spectral index can be approximated to be
\begin{equation}
     n_{s}\simeq 1- \frac{5}{4N}\ ,
\end{equation}
which was found in \cite{Bansal:2024uzr} as well. This quantity practically remains constant for allowed parameters in Fig. \ref{fig:Case 2 region plot}. For $50$-efolds, it can be approximated to be
\begin{equation}\label{ns for case 2 fixed}
    n_s \simeq 0.975. 
\end{equation}
which is in alignment with equation  \ref{ns bound}. Thus, within K\"ahler cone and satisfying the constraints, we can pick a representative value $g_{s}=0.06029$, $\xi=0.5$ and $W_{0}=5$ which spits out the following tensor-to-scalar ratio  
\begin{equation}
    r \simeq 3.5 \times 10^{-8}. 
\end{equation}
It is seen that, all the allowed parameter space forces tensor-to-scalar ratio $r$ to be at least $10^{-8}$ and $n_{s}$ to be around equation  \eqref{ns for case 2 fixed}. Therefore, in this case, constructing a region of validity by imposing bounds equation  \eqref{ns bound} and equation  \eqref{r bound} is redundant unlike case 1 where the region shrinks upon imposing further constraints as evident from Fig. \ref{fig: ns and r region plot 1}. One should note that Case 2 discussed here, even though has a similar potential as loop blow-up inflation \cite{Bansal:2024uzr}, has several interesting differences. As evident from the discussion in this article, our stabilisation mechanism is built with perturbative corrections whereas loop blow-up inflation in \cite{Bansal:2024uzr} operates in LVS. This leads to a different parameter space. By observing our plot Fig. \ref{fig:Case 2 region plot}, it is evident that our tensor-to-scalar ratio is smaller than \cite{Bansal:2024uzr} because we can only go to larger values of $r$ if $W_{0}$ is larger and $\tau_{b}$ is smaller (or subsequently $g_{s}$ is larger) which are restricted by our constraints discussed in Sec. \ref{section:5}. 

\section{Conclusion}
Stabilising moduli and building a successful inflationary model in a way that is consistent in string compactifications remains a daunting task. While non-perturbative corrections to the superpotential can stabilise the K\"ahler moduli, which is flat in leading order, an alternative approach can be taken where perturbative corrections can be used to stabilise the K\"ahler moduli.

In this work, we extend the perturbative stabilisation analysis done in \cite{Cicoli:2024bwq} to include multiple K\"ahler modulus. We explicitly worked out the case of two modulus and the corrections that helped us stabilise the moduli include moduli-redefinitions of both the moduli, $\mathcal{O}(\alpha^{\prime 3})$ correction and 1-loop logarithmic correction at $\mathcal{O}(\alpha^{\prime 3})$ and string-loop corrections at $\mathcal{O}(\alpha^{\prime 2})$. We then turned our attention to focus on possible inflationary scenarios involving the blow-up K\"ahler modulus using perturbative corrections. It turns out that two kinds of inflationary scenarios are possible. The first class of models arise due to the moduli redefinitions of the blow-up modulus. This leads to an inflationary scenario where the inflationary potential has the form $V(\phi)=V_0(1+C_{1} \phi^{2/3})$. In these class of models, we constructed a working region of parameter space where all EFT and cosmological constraints are satisfied leading to a successful inflationary scenario with tensor-to-scalar ratio $r\lesssim 10^{-2}$. Another class of models are due to the string loop corrections which give rise to a potential of the form $V(\phi)=V_0(1+C_{2}\phi^{-2/3})$ which is basically the form of the potential for loop blow-up inflation \cite{Bansal:2024uzr} but differs in stabilisation mechanism which leads to a different region of validity. In these class of models, one can also satisfy both EFT and inflationary constraints with tensor-to-scalar ratio $r\lesssim 10^{-8}$. In the following table, we try to highlight the distinctive features of the two models. 

\begin{table}[h!]
\centering
\begin{tabular}{c|c|c|c||c|c}
Cases & $\phi_*$ & $ g_s  $ & $W_{0}$ & $n_s$ & $r$  \\
\hline
Case 1 & $0.341$ & $0.07077 $ & $7$ & $0.9791$ &  $\simeq 0.028$ \\  \hline
Case 2 & $0.558$ & $0.0603$ & $3$ & $0.975$ &  $\simeq 1.3 \times 10^{-8}$ \\  \hline
\end{tabular}
\caption{Benchmark examples which satisfy all constraints for $N=50$ e-folds.}
\label{tab:Numeric_example}
\end{table}

We end this discussion by pointing out some possible directions. The immediate direction that can be taken is to study the reheating mechanism in these models \cite{Chakraborty:2026hob} and also see the possibility of embedding standard model as done, for example in \cite{Cicoli:2021dhg}. Furthermore, the inflationary scenario arising from the redefinition of the moduli, has not been studied in the context of LVS. This would lead to a different region of parameter space compared to ours. Whether these models also include the possibility of primordial black-hole formations is an exciting direction that we postpone for future work. It is quite an interesting direction to explore, as perturbative corrections was utilised when studying PBH formation in \cite{Cicoli:2018asa}, and one can test to see if our models can be rich enough to account for PBH-formation. The discussion on fibred CY was beyond the scope of this work and an interesting question to ask is if moduli redefinitions of the fibre can lead to a new version of original fibre-inflation \cite{Cicoli:2008gp}. 
\section*{Acknowledgements}
We thank Michele Cicoli, Mahbub Majumdar and Gonzalo Villa for comments on the draft. ARK is supported by Czech Science Foundation GAČR grant ``Dualities and higher derivatives" (GA23-06498S). ARK thanks Luca Brunelli for helpful discussions on loop-blow up inflation, Sitima Moeen, Sayeda Jahan and Saif Ar Rasul for numerous discussions on moduli stabilisation and Christopher Hughes for helpful suggestions on this work. MSJS would like to thank Vladimir Rosenhaus for his support. The work of MSJS is supported through a fellowship from the ITS, funded by a Simons grant and NSF grant 2209116. MH is supported by the Research Seed Grant Initiative 2025 (RSGI) BRACURSGI25014 of BRAC University. MH would like to thank Francisco Pedro for the help he has provided in elucidating topics in LVS, Fibre Inflation, etc. MSJS and ARK are grateful to the ICTP PWF: Bangladesh Jamal Nazrul Islam Winter School 2025 organised by Brac University for facilitating discussions. 
\appendix
\section{Known Perturbative Corrections} \label{Known Corrections}
Higher derivative corrections, arising in string theory due to expansion in $\alpha^{\prime}$ or square-root of the string length, give certain corrections to Einstein's theory. Understanding corrections to Einstein gravity is tantamount for studying cosmology, since it is best studied today using said theory. In this paper, we dealt with type IIB string theory and thus we specifically fixated on $\alpha^{\prime}$ and string loop expansions in type IIB context. The $\alpha^{\prime}$ expansions are essentially higher derivative corrections and loop expansions are expansions in the string coupling $g_{s}=e^{-\phi}$. We denote the action with perturbative corrections as follows
\begin{equation}\label{eq:Perturbative Expansion}
S_{10}=\sum_{m, n=0}^{\infty}\left(\alpha^{\prime}\right)^m g_s^n S_{10}^{(m, n)}\ ,
\end{equation}
where, $m$ and $n$ dictate expansions in $\alpha^{\prime}$ and $g_{s}$. Following the conventions in \cite{Burgess:2020qsc}, we can write $S_{10}^{(m, n)}$ as
\begin{equation}
S_{10}^{(m, n)} \equiv S_{10}^{(p, r, n)} \propto \int \sqrt{-g}\left(\frac{1}{\operatorname{Im} \tau}\right)^{(2 n-p+r+1) / 2}\left(g^{\circ \circ} R_{\circ \circ \circ}^{\circ}\right)^p\left[g^{\circ \circ} g^{\circ \circ} g^{\circ \circ} G_{\circ \circ \circ} G_{\circ \circ \circ}\right]^r+\ldots \ ,
\end{equation}
where $m=p+r-1$ and $\circ$ denotes the index structure. The index $m$ is separated into $p$ and $r$ for keeping track  of powers of the Riemann tensor $R$ and 3-form flux $G_{3}$. The leading tree-level $\alpha^{\prime}$ contribution to the type II superstring occurs at $\alpha^{\prime 3}$ and in our notation it would correspond to setting $m=3, n=0$. The $(\text{Rim})^4$ contribution would thus be $m=3,  p=4, n=0, r=0$ in this notation. 

These $\alpha^{\prime}$ corrections are typically computed using scattering amplitudes and these turn out to be a rather tedious task. Since string theory has the resounding capability of aiding us with a plethora of dualities, it is of great utility to utilise such dualities to compute higher derivative corrections \cite{Wulff:2021fhr,Wulff:2024ips,Wulff:2024mgu,Aggarwal:2025lxf}.\footnote{Many manifestly duality invariant approach has been developed where the most well known example is Double-Field theory \cite{Lescano:2021lup,Lunin:2024vsx,Hronek:2022dyr}. However, as first noted in \cite{Hronek:2020xxi} and further solidified in \cite{Hsia:2024kpi}, these manifestly T-duality invariant approach does not seem to work at $\alpha^{\prime 3}$. However, this does not at all rule out the power of using string dualities to compute higher derivative corrections as the obstruction is there for only manifestly T-duality invariant constructions.} 

\subsection{\texorpdfstring{BBHL $\alpha^{\prime 3}$ Correction}{BBHL alpha'^3 Correction}}
The leading correction to the K\"ahler potential comes from the $\mathcal{O}(\alpha^{\prime 3})$ and $\text{(Riemann)}^{4}$ term which can be captured by a shift $\xi$ in the volume $\mathcal{V}$ \cite{Becker:2002nn}. From the perturbative expansion of the action equation  \eqref{eq:Perturbative Expansion}, we see that this corresponds to  $m=3, p=4, n=0, r=0$. We can derive the equation of motion for the dilaton to order $\alpha^{\prime 3}$ from the 10d action:
 \begin{equation}
    S=\int_{M_{10}} d^{10} x \sqrt{-g_{10}} e^{-2 \phi}\left(R+4(\partial \phi)^2+\alpha^{\prime 3} J_0\right)\ ,
\end{equation}
which has a solution $\phi=\phi_0+\zeta(3) Q / 16$. Here, $Q$ is defined through the 6d Euler integrand $\chi=\int_{C Y_3} d^6 x \sqrt{g} Q$. Here, we are ignoring the fluxes and $J_0$ here is the $\text{(Riemann)}^{4}$ term which is written in terms of the correct index structure as \cite{Wulff:2021fhr,Wulff:2024mgu}
\begin{equation}
J_{0}= \frac{\zeta(3)}{3 \cdot 2^{11}} \left( t_{8} t_{8} R^{4}+\frac{1}{4} \varepsilon_{8} \varepsilon_{8} R^{4} \right)\ .
\end{equation}
When we compactify with 6 compact dimensions, the volume along with $\alpha^{\prime}$ correction shifts as follows:
\begin{equation}
    \mathcal{V} \rightarrow \mathcal{V}+\frac{\hat{\xi}}{2} \equiv \mathcal{V}+\frac{\xi}{2}\left(\frac{S+\overline{S}}{2}\right)^{3 / 2} \equiv \mathcal{V}+\frac{\xi}{2 g_s^{3 / 2}}\ ,
\end{equation}
where,
\begin{equation}
    \hat{\xi}=\frac{\xi}{g_s^{3 / 2}}=\frac{\zeta(3) \chi(X)}{(2 \pi)^3 g_{s}^{3 / 2}} \ .
\end{equation}
Thus, our K\"ahler potential with $\alpha^{\prime 3}$ correction is
\begin{equation}
        K=-2 \ln{\left(\mathcal{V}+ \frac{\xi}{2 g_s^{3 /2}}\right)}\ .
\end{equation}
The above form is useful when we have multiple moduli in our setup. For one K\"ahler modulus with volume $\mathcal{V}=\tau^{3/2}$, we can write
\begin{equation}\label{eq:kahler-bbhl}
    K\simeq -3\ln\left( {\tau}+\frac{\xi s^{3/2}}{3\tau^{1/2}}\right)\, .
\end{equation}
\subsection{Logarithmic Corrections}
One loop at $\mathcal{O}(\alpha^{\prime 3},g_{s}^{2})$ correction which is $m=3, n=2$ considering was found to have a logarithmic structure and used for moduli stabilisation in \cite{Antoniadis:1997eg,Antoniadis:2018hqy,Antoniadis:2019rkh}. Including this $R^{4}$ term in the action,
\begin{equation}
    S=\int_{M_{10}} d^{10} x \sqrt{-g_{10}} e^{-2 \phi}\left(R+\alpha^{\prime 3} f(S,\overline{S})J_0\right)\ .
\end{equation}
where $J_{0}$ here is the $\text{(Riemann)}^{4}$ term which is written in terms of the correct index structure as \cite{Wulff:2021fhr,Wulff:2024mgu}
\begin{equation}
J_{0}= \frac{\zeta(3)}{3 \cdot 2^{11}} \left( t_{8} t_{8} R^{4}+\frac{1}{4} \varepsilon_{8} \varepsilon_{8} R^{4} \right) .
\end{equation}
where $f(S,\overline{S})$ is a function of the dilaton and can be expanded as a non-holomorphic Eisenstein series
\begin{equation}
f(S,\overline{S})= 2 \zeta[3]\left(\frac{S-\overline{S}}{2 i}\right)^{3 / 2}+4\left(\frac{S-\overline{S}}{2 i}\right)^{-1 / 2} \zeta[2]+\left(\frac{S-\overline{S}}{2 i}\right)^{1 / 2} \mathcal{O}\left(e^{-2 \pi s}\right).
\end{equation}
Thus, the leading term has coefficients matching the BBHL piece and the loop term will have a coefficient proportional to $s^{-1/2} \zeta[2]$. We are keeping track of this because they will be important for our subsequent analysis regarding moduli stabilisation and inflation. Incorporating the tree-level $R^4$ term and one-loop term in the 10D type IIB action
\begin{equation}
\mathcal{S}_{10D}^{IIB} \simeq \frac{1}{(2 \pi)^7 \alpha^{\prime 4}} \int_{M_{10}} e^{-2 \phi} R-\frac{6}{(2 \pi)^7 \alpha^{\prime}} \int_{M_{10}}\left(-2 \zeta(3) e^{-2 \phi}-4 \zeta(2)\right) \varepsilon_8 \varepsilon_8 R^4.
\end{equation}
Here, $R$ is the 10d Ricci scalar and $\phi$ is the 10d dilaton. Compactifying on $\text{CY}_3$, we get a tree-level and one loop generated correction to the Einstein-Hilbert action
\begin{equation}
\mathcal{S}_{\text{grav}} = \frac{1}{(2 \pi)^{7} \alpha^{\prime 4}} \int_{M_{4} \times \text{CY}_{3}} e^{-2\phi} R + \frac{\chi}{(2 \pi)^{4} \alpha^{\prime}} \int_{M_{4}}\left(2 \zeta(3) e^{-2\phi_{10}} + 4 \zeta(2)\right) R_{(4)}.
\end{equation}
Here, $M_4$ are the four non-compact dimensions and the Euler character depends on three powers of R as follows:
\begin{equation}
\chi=\frac{3}{4 \pi^3} \int_{\text{CY}_3} R \wedge R \wedge R.
\end{equation}
Interestingly, this Einstein-Hilbert is only possible if we consider 6 compact dimensions. Using $\int d^6 y \sqrt{-g} \sim \mathcal{V}$, we can write the 4 dimensional action in string frame roughly as
\begin{equation}
 \mathcal{S}_{\text{grav}}^{4D} \sim \int_{M_4}\left[\mathcal{V} e^{-2 \phi}+\chi\left(2 \zeta(3) e^{-2 \phi} \mp 4 \zeta[2] \right)\right] R_{(4)}.
\end{equation}
Thus, the BBHL piece and the one loop piece appear as a shift in the volume of the Calabi-Yau. There are two possible corrections arising at that order: one due to the $SL(2,Z)$ completion and another one from the logarithmic correction. From this analysis, \cite{Antoniadis:2018hqy} computed a non-zero contribution at 1 loop from three graviton scattering  which is proportional to the logarithm of the volume of the Calabi-Yau. Including this correction alongside BBHL, the K\"ahler potential stands:
\begin{equation}
    K=-2\ln{\left(\mathcal{V} + \frac{\xi s^{3/2}}{2} + \sigma s^{-1/2}- D s^{-1/2}\ln{\mathcal{V}}\right)}. 
\end{equation}
Here, $\sigma$ is the $SL(2,Z)$ completion piece and D is the coefficient of the one-loop $R^4$ logarithmic corrections. The logarithmic term is proportional to the BBHL piece as $D s^{-1/2} \sim \xi s^{3/2}$ with of course a relative sign difference and the tension of the D7 brane. Thus, $\frac{\xi}{D}= \frac{\zeta(3)}{3 \zeta(2) T_7} \equiv c$ which is used in the main text with the D7 brane tension being $2\pi$. In case of a Calabi-yau with a single K\"ahler modulus, the K\"ahler potential is
\begin{equation} 
    K=-3\ln{\left(\tau+\frac{\xi s^{3/2}}{3\tau^{1/2}}+\left(\sigma s^{-1/2} - Ds^{-1/2}\ln (\tau) \right)/\tau^{1/2}\right)}\,.
\end{equation}

Note that the term proportional to $\sigma$ is subleading than the term proportional to $D$ due to a logarithmic enhancement of the term D.

\subsection{Perturbative String Loop Corrections}
String one-loop loop corrections at $\alpha^{\prime 2}$, which in our perturbative expansion of equation  \eqref{eq:Perturbative Expansion} would correspond to $m=2, n=2$, were computed in \cite{Berg:2005ja,Berg:2007wt} for toroidal orientifolds. They computed two corrections which go as:
\begin{equation}
\begin{aligned}
K_{\text{loops}} \propto  & \sum_{i=1}^3 \frac{\mathcal{A}_i^{(KK)}(U, \overline{U})}{4 \tau_i s}+\sum_{i \neq j \neq k}^3 \frac{\mathcal{B}_k^{(W)}(U, \overline{U})}{4 \tau_i \tau_j} \ .
\end{aligned}
\end{equation}
Here, $U$ denotes the complex structure moduli, $s$ is the dilaton, and the two coefficients are both functions of the complex structure moduli. The generalisation to the case of Calabi-Yau was conjectured to be \cite{Cicoli:2007xp}
\begin{equation} \label{KK correction}
\delta K_{\left(g_s\right)}^{K K} \sim \frac{ \sum_i A_i t^i}{s \mathcal{V}}\ ,
\end{equation}
where we have redefined all the constant terms as $A$. The winding correction was conjectured for the case of Calabi-yau to be \cite{Cicoli:2007xp}

\begin{equation}
\delta K_{\left(g_s\right)}^W \sim \sum_{i=1}^{h_{1,1}} \frac{\mathcal{B}_i^W(U, \overline{U})}{\left(a_{i l} t^l\right) \mathcal{V}}\ .
\end{equation}

An interesting observation is that, the KK-corrections equation  \eqref{KK correction} would be more dominant the tree-level $\alpha^{\prime}{}^3$ correction which was used for moduli stabilisation in LVS \cite{Balasubramanian:2005zx}. However, as proved in \cite{Cicoli:2007xp}, there is an extended no-scale structure which magically cancels the term proportional to $A$ in the F-term scalar potential. In the case of two K\"ahler modulus where one of them corresponds to a blow-up, the corrections in terms of 4-cycles can be written as

\begin{equation}
    K_{g_s}^{(KK)} = \sum_{i=1}^{2} \frac{A'_i t^i}{s \mathcal{V}} \sim \frac{A_b \tau_b^{1/2}}{s \mathcal{V}} + \frac{A_s \tau_s^{1/2}}{s\mathcal{V}} \ ,
\end{equation}

and

\begin{equation}
     K_{g_s}^{(W)} = \sum_{i=1}^{2} \frac{B'_i }{s t^i \mathcal{V}} \sim \frac{B_b}{s \tau_b^{1/2} \mathcal{V}} + \frac{B_s}{s \tau_s^{1/2} \mathcal{V}} \ . 
\end{equation}

\subsection{Moduli Redefinitions}
Apart from the loop and $\alpha^{\prime}$ corrections, field redefinitions can occur at one-loop which can modify the K\"ahler potential. It is well known that in presence of D-branes, 1-loop moduli redefinitions does indeed occur \cite{Antoniadis:1999ge}. This idea stems from the running of gauge couplings in string theory \cite{Dixon:1990pc,Conlon:2009xf,Conlon:2009qa,Weissenbacher:2019mef, Weissenbacher:2020cyf, Klaewer:2020lfg} and at 1-loop the moduli fields gets redefined. Since the complex structure and the dilaton has been fixed at their supersymmetric minima by tree-level ingredients, we do not bother studying their effects. The moduli redefinitions in the K\"ahler sector seems to be the most interesting one as evident from the discussion in the bulk of the paper. It is important for both inflationary analysis and late time dS constructions. In the presence of field theory living on D7-branes, the respective modulus is modified by 1-loop $\beta$-function times the logarithm of the volume 
\begin{equation}
    \tau_{\text {new }}=\tau_{\text {old }}-\alpha \ln \mathcal{V}\ ,
\end{equation}
where, $\alpha$ encodes the 1-loop $\beta$-function coefficient along with some order one numbers. The K\"ahler potential with volume given by equation  \eqref{Volume CY general} is then
\begin{equation}
     K\simeq-2 \ln \left(\frac{\left(\tau _b\right){}^{3/2}-\lambda_s \left(\tau_s\right)^{3/2}}{\gamma }\right). 
\end{equation}
In case of redefinition occurring for both the K\"ahler modulus as exploited in the article, the K\"ahler potential is then
\begin{equation}
     K\simeq-2 \ln \left(\frac{\left(\tau _b-\alpha  \ln \left(\tau _b\right)\right){}^{3/2}-\lambda_s \left(\tau _s+\beta\left(\tau _b\right)\right)^{3/2}}{\gamma }\right). 
\end{equation}
In the case of a single K\"ahler modulus, \cite{Cicoli:2024bwq} used moduli redefinitions to stabilise the modulus and obtain dS vacua and the a redefinition of the volume modulus is supported by the analysis of the redefinition of the big-cycle in \cite{Weissenbacher:2019mef, Weissenbacher:2020cyf, Klaewer:2020lfg}. Furthermore, \cite{Conlon:2010ji} studied the effect of such corrections in the context of LVS where they studied moduli redefinitions of the blow-up mode \cite{Dixon:1990pc,Conlon:2009xf,Conlon:2009qa}.  
\bibliography{biblio}
\bibliographystyle{JHEP}

\end{document}